\documentclass[preprint,longtitle,sort&compress,times,numafflabel]{elsarticle}

\usepackage{amssymb}
\usepackage{amsmath}
\usepackage{lineno}
\usepackage{float}
\usepackage{cuted}
\usepackage{adjustbox}
\usepackage[utf8]{inputenc}
\usepackage[T1]{fontenc}
\usepackage{multirow}

\newcommand\qperp{q_\perp}
\newcommand\Pperp{P_\perp}

\newcommand{\gev}{\mathrm{GeV}}

\newcommand{\ptjet}{\ensuremath{p_{\mathrm{T}}^{\mathrm{jet}}}}

\journal{Physics Letters B}

\begin{document}

\begin{frontmatter}

\date{DESY-24-200, December 2024}

\title{Machine Learning-Assisted Measurement of Lepton-Jet Azimuthal Angular Asymmetries in Deep-Inelastic Scattering at HERA}

\author[47]{V.~Andreev}
\author[31]{M.~Arratia}
\author[43]{A.~Baghdasaryan}
\author[8]{A.~Baty}
\author[37]{K.~Begzsuren}
\author[15]{A.~Bolz}
\author[27]{V.~Boudry}
\author[14]{G.~Brandt}
\author[11]{D.~Britzger}
\author[6]{A.~Buniatyan}
\author[47]{L.~Bystritskaya}
\author[15]{A.J.~Campbell}
\author[44]{K.B.~Cantun~Avila}
\author[25]{K.~Cerny}
\author[11]{V.~Chekelian}
\author[33]{Z.~Chen}
\author[44]{J.G.~Contreras}
\author[29]{J.~Cvach}
\author[21]{J.B.~Dainton}
\author[42]{K.~Daum}
\author[35,39]{A.~Deshpande}
\author[23]{C.~Diaconu}
\author[35]{A.~Drees}
\author[15]{G.~Eckerlin}
\author[40]{S.~Egli}
\author[15]{E.~Elsen}
\author[3]{L.~Favart}
\author[47]{A.~Fedotov}
\author[13]{J.~Feltesse}
\author[15]{M.~Fleischer}
\author[47]{A.~Fomenko}
\author[35]{C.~Gal}
\author[15]{J.~Gayler}
\author[19]{L.~Goerlich}
\author[15]{N.~Gogitidze}
\author[47]{M.~Gouzevitch}
\author[45]{C.~Grab}
\author[21]{T.~Greenshaw}
\author[11]{G.~Grindhammer}
\author[15]{D.~Haidt}
\author[20]{R.C.W.~Henderson}
\author[11]{J.~Hessler}
\author[29]{J.~Hladký}
\author[23]{D.~Hoffmann}
\author[40]{R.~Horisberger}
\author[46]{T.~Hreus}
\author[16]{F.~Huber}
\author[4]{P.M.~Jacobs}
\author[26]{M.~Jacquet}
\author[3]{T.~Janssen}
\author[41]{A.W.~Jung}
\author[15]{J.~Katzy}
\author[11]{C.~Kiesling}
\author[21]{M.~Klein \cormark[deceased]}
\cortext[deceased]{deceased}
\author[15]{C.~Kleinwort}
\author[18]{H.T.~Klest}
\author[15]{R.~Kogler}
\author[21]{P.~Kostka}
\author[21]{J.~Kretzschmar}
\author[15]{D.~Krücker}
\author[15]{K.~Krüger}
\author[22]{M.P.J.~Landon}
\author[15]{W.~Lange}
\author[39]{P.~Laycock}
\author[36]{S.H.~Lee}
\author[15]{S.~Levonian}
\author[17]{W.~Li}
\author[17]{J.~Lin}
\author[15]{K.~Lipka}
\author[15]{B.~List}
\author[15]{J.~List}
\author[11]{B.~Lobodzinski}
\author[31]{O.R.~Long}
\author[47]{E.~Malinovski}
\author[1]{H.-U.~Martyn}
\author[21]{S.J.~Maxfield}
\author[21]{A.~Mehta}
\author[15]{A.B.~Meyer}
\author[15]{J.~Meyer}
\author[19]{S.~Mikocki}
\author[4]{V.M.~Mikuni}
\author[24]{M.M.~Mondal}
\author[46]{K.~M\"uller}
\author[4]{B.~Nachman}
\author[15]{Th.~Naumann}
\author[6]{P.R.~Newman}
\author[15]{C.~Niebuhr}
\author[19]{G.~Nowak}
\author[15]{J.E.~Olsson}
\author[47]{D.~Ozerov}
\author[35]{S.~Park}
\author[26]{C.~Pascaud}
\author[21]{G.D.~Patel}
\author[12]{E.~Perez}
\author[34]{A.~Petrukhin}
\author[28]{I.~Picuric}
\author[15]{D.~Pitzl}
\author[16]{V.~Radescu}
\author[28]{N.~Raicevic}
\author[37]{T.~Ravdandorj}
\author[12]{D.~Reichelt}
\author[29]{P.~Reimer}
\author[22]{E.~Rizvi}
\author[46]{P.~Robmann}
\author[3]{R.~Roosen}
\author[47]{A.~Rostovtsev}
\author[7]{M.~Rotaru}
\author[9]{D.P.C.~Sankey}
\author[16]{M.~Sauter}
\author[23,2]{E.~Sauvan}
\author[15]{S.~Schmitt \cormark[corresponding]}
\cortext[corresponding]{corresponding author}
\author[35]{B.A.~Schmookler}
\author[5]{G.~Schnell}
\author[13]{L.~Schoeffel}
\author[16]{A.~Schöning}
\author[15]{F.~Sefkow}
\author[11]{S.~Shushkevich}
\author[15]{Y.~Soloviev}
\author[19]{P.~Sopicki}
\author[15]{D.~South}
\author[27]{A.~Specka}
\author[15]{M.~Steder}
\author[32]{B.~Stella}
\author[46]{U.~Straumann}
\author[35]{C.~Sun}
\author[30]{T.~Sykora}
\author[6]{P.D.~Thompson}
\author[4]{F.~Torales~Acosta}
\author[22]{D.~Traynor}
\author[37,38]{B.~Tseepeldorj}
\author[39]{Z.~Tu}
\author[35]{G.~Tustin}
\author[30]{A.~Valkárová}
\author[23]{C.~Vallée}
\author[3]{P.~Van~Mechelen}
\author[10]{D.~Wegener}
\author[15]{E.~W\"unsch}
\author[30]{J.~Žáček}
\author[33]{J.~Zhang}
\author[26]{Z.~Zhang}
\author[30]{R.~Žlebčík}
\author[43]{H.~Zohrabyan}
\author[26]{F.~Zomer}
\affiliation[1]{organisation={I. Physikalisches Institut der RWTH, Aachen, Germany
}}
\affiliation[2]{organisation={LAPP, Université de Savoie, CNRS/IN2P3, Annecy-le-Vieux, France
}}
\affiliation[3]{organisation={Inter-University Institute for High Energies ULB-VUB, Brussels and Universiteit Antwerpen, Antwerp, Belgium
}}
\affiliation[4]{organisation={Lawrence Berkeley National Laboratory, Berkeley, CA 94720, USA
}}
\affiliation[5]{organisation={Department of Physics, University of the Basque Country UPV/EHU, 48080 Bilbao, Spain
}}
\affiliation[6]{organisation={School of Physics and Astronomy, University of Birmingham, Birmingham, United Kingdom
}}
\affiliation[7]{organisation={Horia Hulubei National Institute for R\&D in Physics and Nuclear Engineering (IFIN-HH) , Bucharest, Romania
}}
\affiliation[8]{organisation={University of Illinois, Chicago, IL 60607, USA
}}
\affiliation[9]{organisation={STFC, Rutherford Appleton Laboratory, Didcot, Oxfordshire, United Kingdom
}}
\affiliation[10]{organisation={Institut für Physik, TU Dortmund, Dortmund, Germany
}}
\affiliation[11]{organisation={Max-Planck-Institut für Physik, Garching, Germany
}}
\affiliation[12]{organisation={CERN, Geneva, Switzerland
}}
\affiliation[13]{organisation={IRFU, CEA, Université Paris-Saclay, Gif-sur-Yvette, France
}}
\affiliation[14]{organisation={II. Physikalisches Institut, Universität Göttingen, Göttingen, Germany
}}
\affiliation[15]{organisation={Deutsches Elektronen-Synchrotron DESY, Hamburg and Zeuthen, Germany
}}
\affiliation[16]{organisation={Physikalisches Institut, Universität Heidelberg, Heidelberg, Germany
}}
\affiliation[17]{organisation={Rice University, Houston, TX 77005-1827, USA
}}
\affiliation[18]{organisation={Argonne National Laboratory, Lemont, IL 60439, USA
}}
\affiliation[19]{organisation={Institute of Nuclear Physics Polish Academy of Sciences, Krakow, Poland
}}
\affiliation[20]{organisation={Department of Physics, University of Lancaster, Lancaster, United Kingdom
}}
\affiliation[21]{organisation={Department of Physics, University of Liverpool, Liverpool, United Kingdom
}}
\affiliation[22]{organisation={School of Physics and Astronomy, Queen Mary, University of London, London, United Kingdom
}}
\affiliation[23]{organisation={Aix Marseille Univ, CNRS/IN2P3, CPPM, Marseille, France
}}
\affiliation[24]{organisation={National Institute of Science Education and Research, Jatni, Odisha, India
}}
\affiliation[25]{organisation={Joint Laboratory of Optics, Palacký University, Olomouc, Czech Republic
}}
\affiliation[26]{organisation={IJCLab, Université Paris-Saclay, CNRS/IN2P3, Orsay, France
}}
\affiliation[27]{organisation={LLR, Ecole Polytechnique, CNRS/IN2P3, Palaiseau, France
}}
\affiliation[28]{organisation={Faculty of Science, University of Montenegro, Podgorica, Montenegro
}}
\affiliation[29]{organisation={Institute of Physics, Academy of Sciences of the Czech Republic, Praha, Czech Republic
}}
\affiliation[30]{organisation={Faculty of Mathematics and Physics, Charles University, Praha, Czech Republic
}}
\affiliation[31]{organisation={University of California, Riverside, CA 92521, USA
}}
\affiliation[32]{organisation={Dipartimento di Fisica Università di Roma Tre and INFN Roma 3, Roma, Italy
}}
\affiliation[33]{organisation={Shandong University, Shandong, P.R.China
}}
\affiliation[34]{organisation={Fakultät IV - Department für Physik, Universität Siegen, Siegen, Germany
}}
\affiliation[35]{organisation={Stony Brook University, Stony Brook, NY 11794, USA
}}
\affiliation[36]{organisation={University of Tennessee, Knoxville, TN 37996, USA
}}
\affiliation[37]{organisation={Institute of Physics and Technology of the Mongolian Academy of Sciences, Ulaanbaatar, Mongolia
}}
\affiliation[38]{organisation={Ulaanbaatar University, Ulaanbaatar, Mongolia
}}
\affiliation[39]{organisation={Brookhaven National Laboratory, Upton, NY 11973, USA
}}
\affiliation[40]{organisation={Paul Scherrer Institut, Villigen, Switzerland
}}
\affiliation[41]{organisation={Department of Physics and Astronomy, Purdue University, West Lafayette, IN 47907, USA
}}
\affiliation[42]{organisation={Fachbereich C, Universität Wuppertal, Wuppertal, Germany
}}
\affiliation[43]{organisation={Yerevan Physics Institute, Yerevan, Armenia
}}
\affiliation[44]{organisation={Departamento de Fisica Aplicada, CINVESTAV, Mérida, Yucatán, México
}}
\affiliation[45]{organisation={Institut für Teilchenphysik, ETH, Zürich, Switzerland
}}
\affiliation[46]{organisation={Physik-Institut der Universität Zürich, Zürich, Switzerland
}}
\affiliation[47]{organisation={Affiliated with an institute covered by a current or former collaboration agreement with DESY
}}

\begin{abstract}
In deep-inelastic positron-proton scattering, the lepton-jet azimuthal angular asymmetry is measured using data collected with the H1 detector at HERA. When the average transverse momentum of the lepton-jet system, $\lvert \vec{P}_\perp \rvert $, is much larger than the total transverse momentum of the system, $\lvert \vec{q}_\perp \rvert$, the asymmetry between parallel and antiparallel configurations, $\vec{P}_\perp$ and $\vec{q}_\perp$, is expected to be generated by initial and final state soft gluon radiation and can be predicted using perturbation theory.  Quantifying the angular properties of the asymmetry therefore provides an additional test of the strong force. Studying the asymmetry is important for future measurements of intrinsic asymmetries generated by the proton's constituents through Transverse Momentum Dependent (TMD) Parton Distribution Functions (PDFs), where this asymmetry constitutes a dominant background.  Moments of the azimuthal asymmetries are measured using a machine learning method for unfolding that does not require binning. 
\end{abstract}
\end{frontmatter}

\section{Introduction}
In deep-inelastic positron-proton scattering, a high energy lepton\footnote{Lepton is used in this work to refer to positrons or electrons.} scatters of a target proton, producing a high-energy hadronic jet. Due to momentum conservation, the outgoing objects are nearly back-to-back in the transverse plane. Large deviations from this back-to-back configuration can be generated in reactions producing more than two outgoing objects or when one of the outgoing objects undergoes a hard, wide-angle emission. Small deviations can additionally be generated by initial and final state radiation, also from the intrinsic structure of the colliding particles when at least one is a hadron. Sources of these deviations can be modeled by the gluon Wigner distributions~\cite{Hatta:2016dxp, Altinoluk:2015dpi, Zhou:2016rnt, Hagiwara:2017fye, Mantysaari:2019csc, Mantysaari:2019hkq}, intrinsic momenta of polarized gluons within the hadron~\cite{Boer:2010zf,Metz:2011wb,Dumitru:2015gaa,Boer:2017xpy,Boer:2016fqd,Xing:2020hwh,Zhao:2021kae} that are predicted to generate a $\cos(2\phi)$ asymmetry. More recently, asymmetries are predicted to also be sensitive to saturation phenomena, and potentially measurable at the upcoming Electron-Ion Collider (EIC)~\cite{Tong:2022zwp}.   

However, when the total transverse momentum of the jet and scattered lepton, $\lvert \vec{q}_\perp \rvert$, is much smaller than the average transverse momentum of a lepton-jet system, $\lvert \vec{P}_\perp \rvert $, initial and final state soft gluon radiation is predicted to dominate the asymmetry~\cite{Hatta:2021jcd}. This poses a potential problem for other asymmetry measurements that aim to measure the internal structure of the proton, as the soft gluon radiation is not related to the intrinsic structure of the target hadron. A measurement of the asymmetry in this regime would thus provide essential constraints for future explorations of the intrinsic asymmetries.

The presented measurement not only probes this region of phase space where extrinsic contributions to the asymmetry are expected to dominate, but also covers a very large range of $q_\perp$. The measurement is done in an unpolarized system, and serves as a vital testing ground for future measurements at even higher $Q^2$ with polarized hadrons that will further probe the nucleon structure. 

The data used in this study are from deep-inelastic scattering (DIS) events collected with the H1 detector using positron-proton collisions from the HERA collider. Events with a high negative squared momentum transfer  between the lepton and proton, $Q^2$, are examined in the laboratory frame for the imbalance between the scattered lepton and outgoing jet.  Previous measurements of lepton-jet correlations in the laboratory frame using the same dataset explored various kinematic properties~\cite{2108.12376,H1prelim-22-031}.  The measurement in Ref.~\cite{2108.12376} was performed simultaneously in the lepton kinematic properties, the jet transverse\footnote{The present measurement uses a right handed coordinate system defined such that the positive $z$ direction points in the direction of the proton beam and the nominal interaction point is located at $z=0$. The polar angle $\theta$, is defined with respect to this axis. The pseudorapidity is defined as $\eta_{\mathrm{lab}} = -\ln \tan(\theta/2)$. For this work, we are using $\eta = \eta_{\mathrm{lab}}$.} momentum, the jet pseudorapidity $\eta$, the relative transverse lepton-jet momentum imbalance $\qperp/Q$ (where $\qperp$ sum of the lepton and jet transverse momentum), and the angular separation in the transverse plane, $\Delta\phi$.  The results were presented as four binned differential cross section measurements. The goal of the current analysis is to extend the previous results by measuring the moments of the azimuthal asymmetry harmonics as a function of total transverse momentum, $\qperp$.

The previous simultaneous eight-dimensional measurement was enabled by the machine learning-based unfolding method \textsc{MultiFold}~\cite{Andreassen:2019cjw,omnifoldiclr}.  In addition to the lepton-jet studies described above, \textsc{MultiFold} has also been used to measure properties of jet substructure~\cite{H1prelim-22-034,LHCb:2022rky}.  A key feature of \textsc{MultiFold} is that it is unbinned, so the measurement of moments is unaffected by binning artifacts.  As the goal of the present analysis is to measure moments as a function of energy scale, the unbinned nature of \textsc{MultiFold} plays a critical role in achieving a precise result.

This paper is organized as follows. Section~\ref{sec:h1} introduces the H1 detector and the analysis observables.  Then, Sec.~\ref{sec:mc} describes the Monte Carlo simulated datasets used for the analysis. Corrections for detector effects (unfolding) using the \textsc{MultiFold} algorithm are detailed in Sec.~\ref{sec:unfold}. Uncertainty estimation is detailed in Sec.~\ref{sec:uncerts}.  Theoretical predictions using Quantum Chromodynamics (QCD) and experimental results are presented in Sec.~\ref{sec:results} and the paper ends with conclusions and outlook in Sec.~\ref{sec:conclusions}.

\section{Experimental method}\label{sec:h1}
    A full description of the H1 detector can be found elsewhere ~\cite{Abt:1993wz,Andrieu:1993kh,Abt:1996hi,Abt:1996xv,Appuhn:1996na} while the detector components that are most relevant for this measurement are described below.  The main sub-detectors used in this analysis are the inner tracking detectors and the Liquid Argon (LAr) calorimeter, which are both immersed in a magnetic field of 1.16 T provided by a superconducting solenoid. The central tracking system, which covers 15$^{\circ}$ $<\theta<$ 165$^{\circ}$ and the full azimuthal angle, consists of drift and proportional chambers that are complemented with a silicon vertex detector in the range $30^{\circ}<\theta<150^{\circ}$~\cite{Pitzl:2000wz}. It yields a transverse momentum resolution for charged particles of $\sigma_{p_\mathrm{T}}/p_\mathrm{T}$ = 0.2$\%$ $p_\mathrm{T}$/GeV$~\oplus~$1.5$\%$. The LAr calorimeter, which covers $4^{\circ}<\theta< 154^{\circ}$ and full azimuthal angle, consists of an electromagnetic section made of lead absorbers and a hadronic section with steel absorbers; both are highly segmented in the transverse and longitudinal directions. Its energy resolution is $\sigma_{E}/E = 11\%/\sqrt{E/\mathrm{GeV}}$~$\oplus$~$1\%$ for leptons~\cite{Andrieu:1994yn} and $\sigma_{E}/E\approx 50\%/\sqrt{E/\mathrm{GeV}}$~$\oplus$~$3\%$ for charged pions~\cite{Andrieu:1993tz}.  In the backward region ($153^\circ < \theta < 177.5^\circ$), energies are measured with a lead-scintillating fiber calorimeter~\cite{Appuhn:1996na}.
Results are reported using the data recorded by the H1 detector in the years 2006 and 2007 when positrons and protons were collided at energies of 27.6 GeV and 920 GeV, respectively.  The total integrated luminosity of this data sample corresponds to 136 pb$^{-1}$ \cite{H1:2012wor}.

DIS reactions are governed by $Q^2$ and the inelasticity $y$, or equivalently, the longitudinal momentum fraction $x=Q^2/(sy)$.  The $\Sigma$ method~\cite{Bassler:1994uq} is used to reconstruct $Q^{2}$ and $y$ as: 
\begin{align}\label{eq:q2}
Q^{2} &=\frac{E_{e^{'}}^2 \sin^{2}\theta_{e^\prime}}{1-y}\,\\
y &= \frac{\sum_{i\in \mathrm{had}}(E_{i}-p_{i,z})}{\sum_{i\in \mathrm{had}}(E_{i}-p_{i,z})+E_{e^{'}}(1-\cos\theta_{e^{'}})}\,,
\label{kinematicreco}
\end{align}
where $\theta_{e^{\prime}}$ is the polar angle of the scattered lepton and $\sum(E_{i}-p_{i,z})$ is the total difference between the energy and longitudinal momentum of the entire hadronic final state.  Compared to other methods, the $\Sigma$ reconstruction method reduces sensitivity to collinear initial state QED radiation, $e\to e\gamma$, since the beam energies are not included in the calculation. 


Events are triggered by requiring a high energy cluster in the electromagnetic part of the LAr calorimeter. The scattered lepton is identified as the highest transverse momentum LAr cluster matched to a track passing an isolation criteria~\cite{Adloff:2003uh}. Events containing scattered leptons with energy $E_{e'}>11$ GeV are kept for further analysis, resulting in a trigger efficiency higher than 99.5$\%$~\cite{Aaron:2012qi,Andreev:2014wwa}. Backgrounds from additional processes such as cosmic rays, beam-gas interactions, photoproduction, charged-current DIS and Quantum Electrodynamics (QED) Compton processes are rejected after dedicated selection~\cite{Andreev:2014wwa,Andreev:2016tgi}, resulting in negligible background contamination.

The \textsc{FastJet}~3.3.2 package~\cite{Cacciari:2011ma,Cacciari:2005hq} is used to cluster jets in the laboratory frame with
the inclusive $k_{\mathrm{T}}$ algorithm~\cite{Catani:1993hr,Ellis:1993tq} and distance parameter $R = 1$. The inputs for the jet clustering are hadronic final state (HFS) objects with $-1.5<\eta<2.75$.  These objects are built from calorimeter-cell clusters and reconstructed tracks, after removing those associated with the scattered lepton, using an energy flow algorithm~\cite{energyflowthesis,energyflowthesis2,energyflowthesis3}.  Jets with transverse momentum $\ptjet>$ 5 GeV are selected for further analysis. %

Events with $Q^{2}> 100$ GeV$^{2}$, $0.08<y<0.7$, and at least one jet participate in the unbinned unfolding (Sec.~\ref{sec:unfold}). The final measurement is presented using the leading jet in the event within a fiducial volume defined by $Q^2>150$ GeV$^2$, $0.2 < y < 0.7$, $\ptjet>$ 10 GeV, $-1.0<\eta^{\rm jet}<2.5$, and $\qperp / \ptjet \leq 0.3$. 

\begin{figure}[h]
    \centering
    \includegraphics[width=0.7  \textwidth]{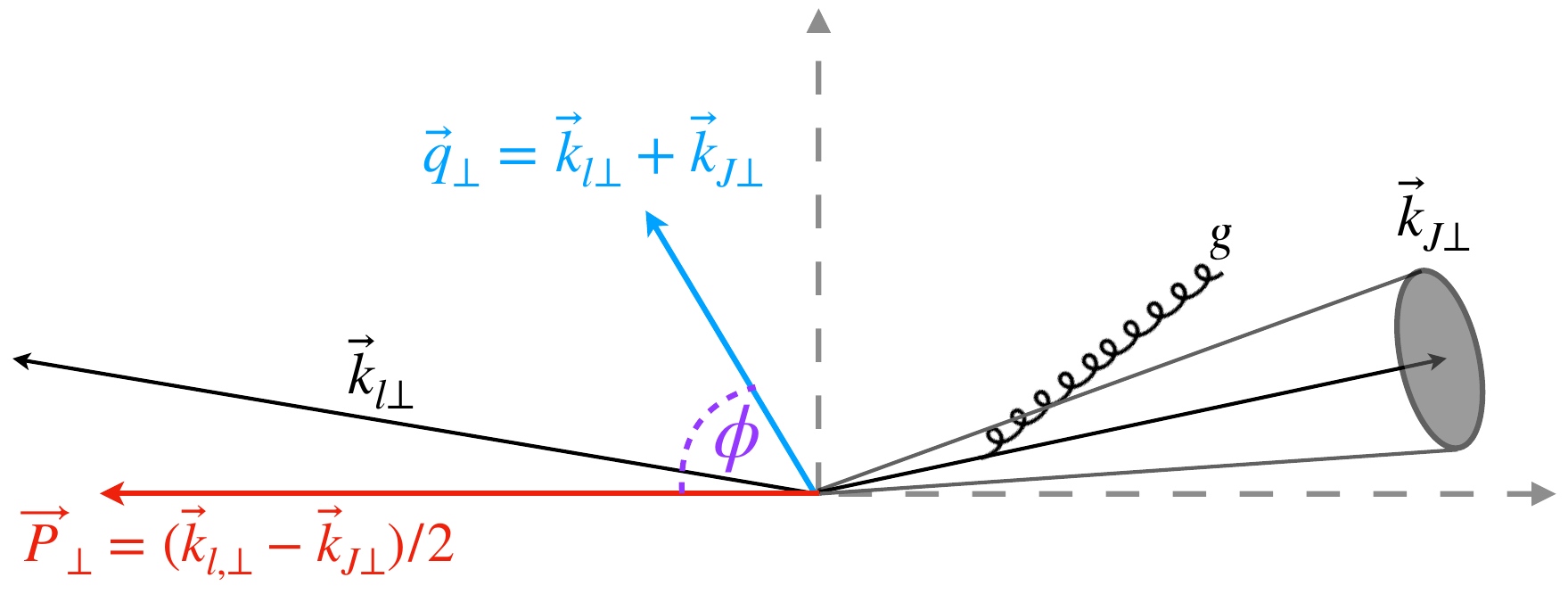
} %
    \caption{Lepton-jet final state in the transverse plane perpendicular to the beam
axis. The lepton-jet total transverse momentum is labeled $\vec{q}_\perp = \vec{k}_{l\perp} + \vec{k}_{\mathrm{J}\perp}$. The average lepton-jet transverse momentum is denoted $\vec{P}_\perp = (\vec{k}_{l\perp} + \vec{k}_{\mathrm{J}\perp})/2$. The angle between the two vectors is designated $\phi$. Soft gluon radiation tends to be collinear to the jet axis, but can result in a measurable lepton-jet momentum asymmetry if it falls outside of the jet radius. Non-zero $\vec{\qperp}$ can be generated by gluon radiation, as indicated in the simple schematic, or by the intrinsic $p_\mathrm{T}$ of the parton.}
    \label{fig:signature}
\end{figure}

This work aims to measure the azimuthal angular asymmetry between the scattered lepton and the leading reconstructed jet - see Fig.~\ref{fig:signature}. This angle is denoted as $\phi$, and is calculated as the azimuthal angle between the total lepton-jet transverse momentum, 
\begin{equation}
\vec{q}_\perp = \vec{k}_{l\perp} + \vec{k}_{\mathrm{J}\perp},
\end{equation}

and the average lepton-jet transverse momentum, 
\begin{equation}
\vec{P}_\perp = (\vec{k}_{l\perp} - \vec{k}_{\mathrm{J}\perp})/2,
\end{equation}

such that $\phi$ is given by:
\begin{equation}
    \label{eq:phi}
    \phi = \cos^{-1} [ (\vec{q}_\perp \cdot \vec{P}_\perp) / (\lvert\vec{q}_\perp \rvert \cdot\lvert \vec{P}_\perp  \rvert)].
\end{equation}

The measurement is presented in terms of three  harmonics of $\phi$: $\cos{(\phi)}, \ \cos{(2\phi)}$, and $\cos{(3\phi)}$. For events where $\Pperp \gg \qperp$, soft gluon radiation can dominate the asymmetry. This is because radiative corrections are predicted to be enhanced by large double logarithms: $\alpha\ln ^2(\Pperp ^2/\qperp ^2)^n$\cite{Hatta:2021jcd,Tong:2022zwp}.
 To facilitate comparisons to perturbative QCD (pQCD) calculations and to satisfy the condition $\Pperp \gg \qperp$, only events with $\qperp / \ptjet \leq 0.3$ are selected.  Note that $\Pperp \approx \ptjet$ after our kinematic selection.



\section{Monte Carlo simulations}\label{sec:mc}
Monte Carlo (MC) simulations are used to correct the data for detector acceptance and resolution effects as well as to compare theoretical predictions with experimental results. The input for the jet clustering at the generator level (``particle level'') are final-state particles with proper lifetime $c\tau > 10$~mm, excluding the scattered lepton. Leading reconstructed jets (reco) are matched to leading generated jets (gen). This matching of reco-level and particle-level jets is required for Step 2 of the Multifold unfolding procedure, discussed in Sec. \ref{sec:unfold}.

Detector acceptance and resolution effects are estimated using the DJANGOH 1.4~\cite{Charchula:1994kf} and RAPGAP 3.1~\cite{Jung:1993gf} simulators. Both generators implement Born level matrix elements for neutral current DIS, boson–gluon fusion, and QCD Compton processes and are interfaced with \textsc{Heracles}~\cite{Spiesberger:237380,Kwiatkowski:1990cx,Kwiatkowski:1990es} for QED radiation.  The CTEQ6L PDF set~\cite{Pumplin:2002vw} and the Lund hadronization model~\cite{Andersson:1983ia} with parameters determined by the ALEPH Collaboration~\cite{Schael:2004ux} are used for the non-perturbative components.  DJANGOH uses the Color Dipole Model as implemented in \textsc{Ariadne}~\cite{Lonnblad:1992tz} for higher order emissions, and RAPGAP uses parton showers in the leading logarithmic approximation. Each of these generators is combined with a detailed simulation of the H1 detector response based on \textsc{Geant}3~\cite{Brun:1987ma} and the generated events are reconstructed in the same way as data.  

Predictions from Pythia~8.3~\cite{Sjostrand:2006za,Sjostrand:2014zea} are used for comparison using the default implementation and one additional parton shower implementations \textsc{Dire} \cite{Hoche:2015sya}. The  NNPDF3.1 PDF set \cite{NNPDF:2017mvq} default Pythia implementation and MMHT14nlo68cl PDF set \cite{Harland-Lang:2014zoa} is used for the \textsc{Dire} implementation. 

\section{Unfolding}
\label{sec:unfold}
The unfolding procedure for this measurement, \textsc{MultiFold}, is the same one used in \cite{PhysRevLett.128.132002}. 
The \textsc{MultiFold} method is an iterative two-step (expectation-maximization) procedure to correct for detector effects. The goal is to infer particle-level data using detector-level data and simulations. The main components of \textsc{MultiFold} are explained in more detail below.

The unfolded phase space will consist of the quantities $(p_x^e, p_y^e, p_z^e, p_\mathrm{T}^{\text{jet}}, \eta^{\text{jet}}, \phi^{\text{jet}}, q_\perp/Q^2, \Delta\phi^{\text{jet}})$ that are represented as the vector $\vec{x}$.  $p_x^e, p_y^e,$ and $p_z^e$ are the $\hat{x},\hat{y}$, and $\hat{z}$ component of the lepton momentum in cartesian coordinates, respectively. $p_\mathrm{T}^{\text{jet}}$ and $\eta^{\text{jet}}$ are the transverse momentum and pseudorapidity of the jet. $\phi^{\text{jet}}$ is the azimuthal angle of the jet, measured in the transverse plane in the lab system, not to be confused with the primary observable of this work, $\phi$, which is taken as the angle between $\vec{P_\perp}$ and $\vec{q}_\perp$. Lastly, $\Delta\phi^{\text{jet}}$ is the azimuthal angle measured between the jet and the lepton in the transverse plane.

The first step of \textsc{MultiFold} uses observables at detector level while the second step operates at particle level.  Let $X_\text{data}=\{\vec{x}_i\}$ be the set of events in data and $X_\text{MC,gen}=\{\vec{x}_\text{gen,$i$}\}$ and $X_\text{MC,reco}=\{\vec{x}_\text{reco,$i$}\}$ be sets of events in simulation with a correspondance between the two sets.  In simulation, there is a set of observables at particle-level and detector-level for each event.  If an event does not pass the particle-level or detector-level event selection, then the observables for that event are assigned a dummy value $\vec{x}=\emptyset$.  Each event $i$ in simulation is also associated with a weight $w_i$ from the MC simulation.  

\textsc{MultiFold} achieves an unbinned unfolding by iteratively reweighting the particle-level events.  Each event $i$ in simulation is given a weight $\nu_i$ and these weights are updated at each iteration.  The final result is the simulated events with weights $\nu_iw_i$.  From these events, one can compute new observables defined on $\vec{x}$ and can construct histograms or other summary statistics.  The \textsc{MultiFold} weights are initialized at $\nu_i=1$, i.e. the prior is the initial MC simulation.

The first step of \textsc{MultiFold} is to train a classifier $f$ to distinguish the weighted simulation at detector-level from the data.  The classifier is trained to maximize the common binary cross entropy:
    
    \begin{align}
        \varepsilon = \sum_{\vec{x}_i\in X_\text{data}}\log(f(\vec{x}_i)) + \sum_{\vec{x}_i\in X_\text{MC,reco}}\nu_i\,w_i\, \log(1-f(\vec{x}_i))\,,
         \end{align}

where both sums only include events that pass the detector-level selection.  For events that pass the detector-level selection, define $\lambda_i=\nu_i\cdot f(\vec{x}_i)/(1-f(\vec{x}_i))$ for $\vec{x}_i\in X_\text{MC,reco}$.  This manipulation of the classifier output is known (see e.g. Refs.~\cite{hastie01statisticallearning,sugiyama_suzuki_kanamori_2012,Rizvi2024}) to produce an estimate of the likelihood ratio between data and simulation. For events that do not pass the detector-level selection, $\lambda_i=\nu_i$.  
 
 The second step of \textsc{MultiFold} is a regularization step.  The weights $\lambda_i$ are insufficient because they are not a proper function of the particle-level phase space.  In other words, a single phase space point $\vec{x}_{\text{gen}}$ can be mapped to different $\vec{x}_{\text{reco}}$ values under the stochastic detector response.  The second step of \textsc{MultiFold} averages the weights $\lambda$ for a fixed particle-level phase space point.  This is accomplished by training a classifier to distinguish the particle-level simulation weighted by $\nu$ from the particle-level simulation weighted by $\lambda$.  The loss function is once again the binary cross entropy:
    \begin{align}
        {\cal L} = \sum_{\vec{x}_i\in X_\text{MC,truth}}\lambda_i\,w_i\, \log(f(\vec{x}_i)) + \nu_i\,w_i\,\log(1-f(\vec{x}_i))\,,
        \end{align}

     where the sum only includes events that pass the particle-level selection. For events that pass the particle-level selection, define $\nu_i=\nu_i\times f(\vec{x}_i)/(1-f(\vec{x}_i))$ for $\vec{x}_i\in X_\text{MC,truth}$.  For events that do not pass the particle-level selection, $\nu_i$ is left unchanged from its previous value.
    
The classifiers for Steps 1 and 2 are parameterized as fully connected deep neural networks.  These networks are implemented in \textsc{TensorFlow}~\cite{tensorflow} and \textsc{Keras}~\cite{keras} and optimized using \textsc{Adam}~\cite{adam}.  The input layer to the neural networks has 8 nodes, corresponding to the 8 dimensions of $\vec{x}$ used for unfolding.  All inputs are scaled so that each dimension of $\vec{x}$ has mean zero and unit standard deviation. Following the input, there are three hidden layers, with 50, 100, and 50 nodes, respectively. Each layer has a rectified linear unit activation function and the network output is a single node with the sigmoid activation function.  None of these hyperparameters were optimized and all other hyperparameters are set to their default values.  In particular, the network biases are all initialized to zero and the weights are initialized using the Glorot uniform distribution~\cite{journals/jmlr/GlorotB10}.  In order to minimize variations from the stochastic nature of the training procedure, 5 networks are trained for each configuration and the final result is taken as the mean over the 10 values per event.

For training, the inputs are partitioned equally into training and validation sets.  This partition is random and redone at each iteration.   Training proceeds for 10,000 epochs with an early stopping mechanism that halts training if the validation loss does not decrease for 10 consecutive epochs. The training in both steps uses a batch size of 4,000 events and a learning rate of $2\times 10^{-6}$. The learning rate is decreased if there is no improvement in the validation loss after 5 epochs, and the training ends if there is no improvement after 10 epochs. The networks are trained using NVIDIA H100 Graphical Processing Units (GPUs).  These GPUs have sufficient memory (24 GB) to simultaneously fit all of the inputs and the model into memory.  The training time for both Step 1 and Step 2 decreases with each iteration since the MC at particle level is reweighted to successively better match the data with each iteration. 

\section{Uncertainties}\label{sec:uncerts}


Systematic uncertainties on the description of the detector are estimated by varying the relevant aspects of the simulation and carrying out the full analysis procedure with the varied simulation set.

As the main observable of this work, $\phi$, is calculated from the outgoing 4-momenta of the scattered lepton and reconstructed jet, their uncertainties (in particular energy scale and azimuthal angle) are carefully considered. The uncertainties on the HFS energy scale are categorized into two classifications: HFS objects within high $p_{\mathrm{T}}$ jets and all other remaining HFS objects. The energy-scale uncertainty in both cases is $\pm1\%$. Both sources of uncertainty are estimated separately \cite{etde_21406988,H1:2014cbm} by varying each HFS energy by $\pm1\%$. An uncertainty of $\pm20$ mrad is assigned to the azimuthal angle determination of HFS objects. The uncertainties on the lepton energy scale ranges from $\pm0.5\%$ to $\pm1\%$ \cite{H1:2011unn,H1:2014cbm}. Uncertainties on the azimuthal angle of the scattered lepton are estimated to be $\pm1$ mrad \cite{H1:2012qti}. For each variation to the simulation, the models for the unfolding are completely retrained and the unfolding procedure is repeated. The full difference in the final observable from the nominal result is taken as the uncertainty for that variation.

QED corrections  accounting for virtual and real higher-order QED effects are taken as an uncertainty and are estimated by comparing the effect on the final observable with and without initial QED radiation with residual differences taken as the uncertainty. 

An additional uncertainty from the unfolding procedure is estimated to cover a possible bias from the generator choice used to perform the unfolding procedure. This is designated as the model bias, and is estimated by the difference in results obtained when performing the unfolding with the RAPGAP or DJANGOH simulations. The model uncertainty is the leading uncertainty for this measurement. After further investigation, this uncertainty arises from the limited detector resolution of the asymmetry angle and total transverse momentum, $\phi$ and $\qperp$ respectively. The resolution is on the order of 1 radian, and impacts how the unfolding converges.

\begin{figure}
    \centering
 \includegraphics[width=1.0\textwidth]{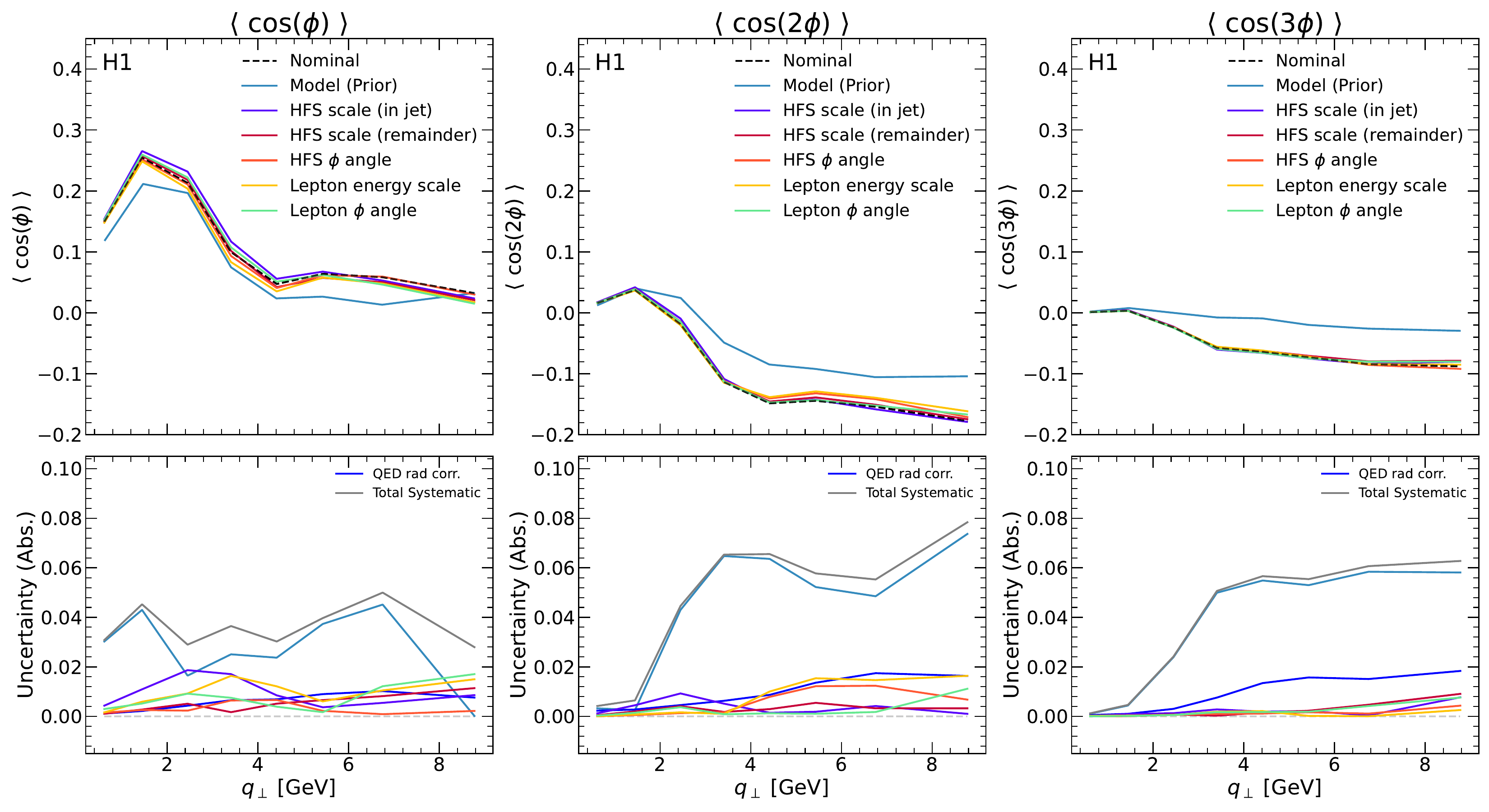
}
    \caption{The systematic uncertainties for this measurement. The nominal results (black dashes) are plotted against the systematic variations taken for estimating each uncertainty. The bottom panels show the absolute uncertainties obtained from each variation, and additionally include the QED uncertainty, as well as the total systematic uncertainty.}
    \label{fig:uncertanties}
    \label{fig:my_label}
\end{figure}

The major sources of uncertainty are plotted in Fig. \ref{fig:uncertanties} as a function of this measurement's independent variable, $\qperp$. The uncertainties are reported as absolute quantities.


The statistical uncertainty of the measurement is estimated using the bootstrap technique~\cite{10.1214/aos/1176344552}. The unfolding procedure is repeated on 100 pseudo datasets, each defined by resampling the original dataset according to a Poisson distribution with $\mu=1$. The number of MC events exceeds the number of data events by nearly two orders of magnitude and therefore the MC statistical uncertainty is negligible compared to the corresponding data uncertainty.  Due to the ensembling procedure described in Sec.~\ref{sec:unfold}, variations from the random nature of the network initialization and training are negligible compared to the statistical uncertainty of the data.

The measurement is of moments of the asymmetry angle distribution, and is normalized by definition. As a result, uncertainties relating to luminosity and triggers do not impact the results.



\section{Results}\label{sec:results}
\begin{figure}[h]
    \centering
    \includegraphics[width=0.98\textwidth]{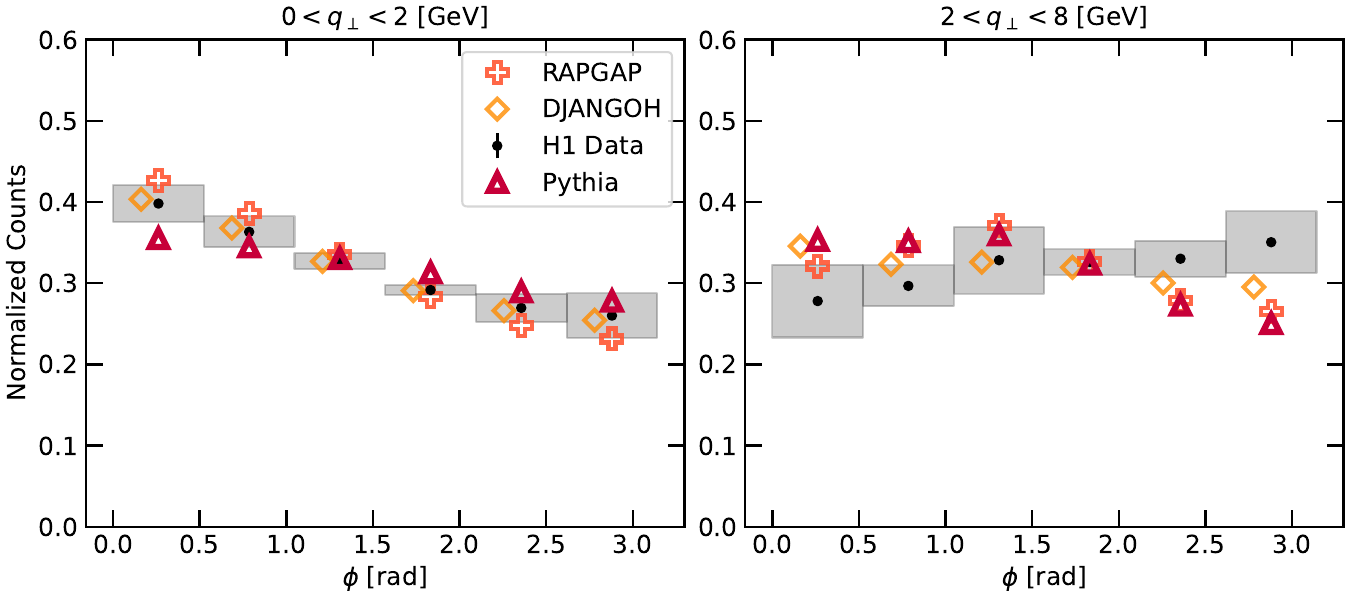
} %
    \caption{Azimuthal angular asymmetry between the lepton and leading jet in two bins of $q_\perp$. Data are shown in black, and compared with RAPGAP and DJANGOH, shown in orange and yellow respectively. Pythia is shown is red triangles. There is good agreement between the data and both simulations in the first $q_\perp$ bin, but large disagreement for higher $q_\perp$.}
    \label{fig:phi_results}
\end{figure}

The unfolded asymmetry angle distributions are shown in Fig. \ref{fig:phi_results}, where data is compared to RAPGAP and DJANGOH Monte-Carlo DIS event generators, as well as Pythia 8.2. The lower $q_\perp$ bin shows a clear cosine-like distribution, expected for the lepton-jet configuration, and in agreement with the prediction that the first harmonic will be the largest. The data are in good agreement with both DIS event generators. They also show good agreement with Pythia, however there is a deviation at the low values of $\phi$. The high $q_\perp$ bin shows a more non-trivial $\phi$ distribution. The data show a slightly positive slope towards higher values of $\phi$, but while data generally maintains good agreement with the event generators, the discrepancies at the tails of the distribution are larger. The asymmetry angle results are summarized in Table \ref{tab:phi}.

\begin{figure}[h]
    \centering
    \includegraphics[width=1\textwidth,trim=5 5 5 5,clip]{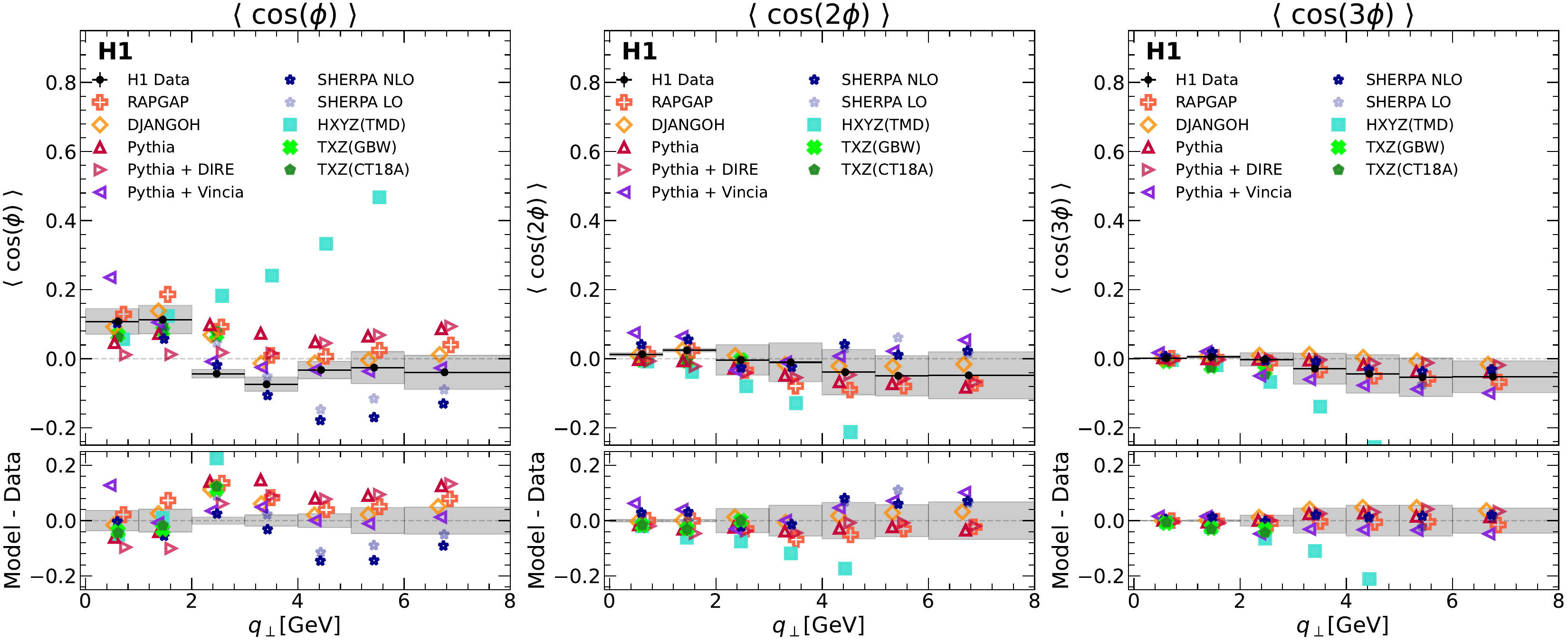
} %
    \caption{Three harmonics of the azimuthal angular asymmetry between the lepton and leading jet as a function of $\qperp$. Predictions from multiple simulations as well as a pQCD calculation are shown for comparison. The absolute difference between data (black circles) and predictions is shown in the bottom panels. The gray band represents the total systematic uncertainty, while the vertical black bars on the data points (often masked by the markers) represent the statistical uncertainty.}
    \label{fig:results}
\end{figure}

The unfolded $\cos(\phi)$, $\cos(2\phi)$, and $\cos(3\phi)$ harmonics of the lepton-jet azimuthal angular asymmetry are shown in Fig. \ref{fig:results} and are compared to various Monte Carlo event generators (Pythia, RAPGAP, DJANGOH and SHERPA). The measured asymmetry is largest for the first harmonic, $\cos(\phi)$, but only for small $\qperp$. The second and third harmonics, $\cos(2\phi)$ and $\cos(3\phi)$, are consistent with no asymmetry within uncertainties. This supports the initial expectation in two key ways: First, that the first harmonic will have the largest asymmetry in the lepton-jet configuration. Second, the asymmetry is high only in low $\qperp$ events. The regime where soft gluon radiation significantly contributes to the lepton-jet asymmetry was estimated to be $\qperp < 3.0$ GeV \cite{Tong:2022zwp}. The harmonics are summarized in Table \ref{tab:main_results}.

The unfolded data are compared with two sets of calculations. The first set is re-calculated from Hatta et al.~\cite{Hatta:2021jcd} for HERA kinematics, and is shown with the label HXYZ (TMD). The soft gluon radiation is predicted to be enhanced by the double logarithm, $(\alpha_s \ln^2 {P_\perp}/{q_\perp})^n$, and the calculations focus on the re-summation of this double logarithm. The predictions were done in the TMD factorization framework. The predictions are shown as blue-green squares in Fig~\ref{fig:results}. The cross section for these predictions is stated to be well behaved as $q_\perp \rightarrow 0$. However, the prediction was made up to a higher $q_\perp$ of 6.0 GeV. 

The asymmetry from soft gluon radiation was also calculated with kinematics relevant for the upcoming electron ion collider experiment in \cite{Tong:2022zwp}, and labeled as ZXT. While the observable is the same, the asymmetry was calculated in the context of probing gluon saturation effects in electron-nucleus collisions. The calculations were re-done for HERA kinematics however, with a jet resolution parameter of 1.0 and they use a similar framework as the calculations done in \cite{Hatta:2021jcd}. One of the calculations expresses the quark distribution in terms of a dipole scattering amplitude. The authors then use a 3-parameter model fit to HERA data labeled as Golec-Biernat and Wusthoff (GBW) model, that aims to describe saturation phenomena at low $Q^2$ \cite{Albacete:2010sy}. The second calculation uses the \textsc{CT18A} PDF that involves a next-to-leading order (NLO) TMD calculation using collinear PDFs \cite{Hou:2019efy}, and does not include saturation phenomena. Importantly, both calculations require small $\qperp$, specifically $\qperp < 3.0$ GeV.  Both calculations agree with the measurement for the second and third harmonics, within the presented uncertainties. Both calculations also agree with the data quite well in the first harmonic, with the exception of the last $\qperp$ bin. This may indicate that the asymmetry arising from soft gluon radiation may fall off earlier in $\qperp$ than expected, as there is good agreement only up to $\qperp < 2.0$ $\gev$ for the first harmonic. This also indicates that the difference between the two calculations (the inclusion of gluon saturation effects) cannot be measured with the current precision at this kinematic range.

 While both sets of predictions in \cite{Tong:2022zwp} and \cite{Hatta:2021jcd} use similar theoretical frameworks, the original calculation from Hatta et al. were done in bins of $q_\perp/P_\perp$, and re-scaled to $q_\perp$. When scaling back to $q_\perp$, a simplifying assumption of $p_\mathrm{t,jet} \approx P_\perp$ was made during the re-scaling process, which may be one contributing factor to the disagreement with data, particularly at high $q_\perp$.

Three Pythia models, namely Pythia \cite{Sjostrand:2006za,Sjostrand:2014zea}, Pythia DIRE \cite{Hoche:2015sya} and Pythia Vincia \cite{Giele:2007di,Giele:2013ema}, are also compared to the data with varying agreement for each harmonic.  Pythia and Pythia + DIRE are very similar for the most part, and show good agreement with the measured data for all $\qperp$ bins for $\cos(2\phi)$, but deviated from the data at low $\qperp$ for $\cos(\phi)$. All three Pythia predictions for $\cos(3\phi)$ agree with data within uncertainties. RAPGAP and DJANGOH agree with the measured data below $\qperp = 2.0$ GeV, similar to the TXZ(GBW) and TXZ(CT18A) calculations. RAPGAP and DJANGOH also agree with the measurement at higher $\qperp$, where the data indicate no asymmetry. Lastly, predictions from SHERPA 3.0.1 \cite{Sherpa:2024mfk} are shown. The predictions include NLO matrix element calculations via the MC@NLO method as implemented in SHERPA \cite{Hoeche:2011fd}. Events are showered with SHERPA's default dipole based parton shower \cite{Schumann:2007mg} and hadronised with a new implementation of SHERPA's cluster hadronization model \cite{Chahal:2022rid}, tuned to data from LEP \cite{Knobbe:2023njd} and HERA \cite{Knobbe:2023ehi}. 
SHERPA NLO refers to the MC@NLO prediction. For comparison, the LO predictions showered with SHERPA are included as well. Both Sherpa predictions underestimate the first harmonic in particular at large $q_\perp$, but agree with the measured data for the second and third harmonic. 

\begin{table}[h]
\centering
\footnotesize
\caption{Mean counts of the lepton-jet asymmetry angle $\phi$, in two $q_\perp$ intervals. The statistical (Stat.) and total (Tot.) uncertainties are given, as well as the seven systematic uncertainties described in the main text.}

\begin{tabular}{|c|c|c|c||c|c|c|c|c|c|c|}

\hline
\multicolumn{11}{|c|}{$0.0 < q_\perp < 2.0$ [GeV]} \\
\hline

$\phi$ [rad.] & count & Stat. & Tot. & Model & HFS(jet) & HFS(other) & HFS($\phi$) & Lepton(E) & Lepton($\phi$) & QED \\
\hline
$[0.0,0.52]$ & 0.398 & 0.002 & 0.023 & 0.017 & 0.003 & 0.001 & 0.014 & 0.001 & 0.001 & 0.002 \\                                $[0.52,1.05]$ & 0.363 & 0.001 & 0.019 & 0.016 & 0.004 & 0.0 & 0.009 & 0.001 & 0.002 & 0.003 \\
$[1.05,1.57]$ & 0.327 & 0.0 & 0.01 & 0.009 & 0.001 & 0.0 & 0.002 & 0.002 & 0.0 & 0.003 \\                                      $[1.57,2.09]$ & 0.292 & 0.001 & 0.006 & 0.004 & 0.002 & 0.0 & 0.004 & 0.0 & 0.001 & 0.002 \\                                   $[2.09,2.62]$ & 0.27 & 0.001 & 0.017 & 0.014 & 0.003 & 0.0 & 0.009 & 0.002 & 0.001 & 0.002 \\
$[2.62,3.14]$ & 0.26 & 0.001 & 0.028 & 0.024 & 0.003 & 0.001 & 0.013 & 0.003 & 0.001 & 0.002 \\                    \hline
\multicolumn{11}{|c|}{$2.0 < q_\perp < 8.0$ [GeV]} \\
\hline
$\phi$ [rad.] & count & Stat. & Tot. & Model & HFS(jet) & HFS(other) & HFS($\phi$) & Lepton(E) & Lepton($\phi$) & QED \\
\hline
$[0.0,0.52]$ & 0.278 & 0.003 & 0.045 & 0.039 & 0.0 & 0.003 & 0.021 & 0.005 & 0.002 & 0.003 \\
$[0.52,1.05]$ & 0.297 & 0.002 & 0.025 & 0.015 & 0.004 & 0.002 & 0.019 & 0.004 & 0.0 & 0.004 \\
$[1.05,1.57]$ & 0.328 & 0.001 & 0.041 & 0.039 & 0.002 & 0.003 & 0.011 & 0.006 & 0.001 & 0.004 \\
$[1.57,2.09]$ & 0.326 & 0.002 & 0.016 & 0.014 & 0.003 & 0.002 & 0.003 & 0.004 & 0.002 & 0.003 \\
$[2.09,2.62]$ & 0.33 & 0.002 & 0.022 & 0.009 & 0.001 & 0.005 & 0.018 & 0.007 & 0.003 & 0.004 \\
$[2.62,3.14]$ & 0.35 & 0.002 & 0.038 & 0.02 & 0.003 & 0.005 & 0.029 & 0.012 & 0.002 & 0.005 \\
\hline

\end{tabular}
\label{tab:phi}
\end{table}

\begin{table}[h]
\centering
\footnotesize
\caption{The measured means of the first three harmonics of the lepton-jet asymmetry angle $\phi$, in bins of $q_\perp$. The statistical (Stat.) and total (Tot.) uncertainties are given, as well as the seven systematic uncertainties described in the main text.}

\begin{tabular}{|c|c|c|c||c|c|c|c|c|c|c|}

\hline
$q_\perp$ [GeV] & $\langle \cos(\phi) \rangle$ & Stat. & Tot. & Model & HFS(jet) & HFS(other) & HFS($\phi$) & Lepton(E) & Lepton($\phi$) & QED \\
\hline

$[0.0, 1.0]$ & $0.10$8 & 0.001 & 0.037 & 0.037 & 0.003 & 0.001 & 0.001 & 0.001 & 0.001 & 0.0 \\
$[1.0, 2.0]$ & 0.113 & 0.003 & 0.041 & 0.041 & 0.004 & 0.001 & 0.001 & 0.0 & 0.003 & 0.003 \\
$[2.0, 3.0]$ & $-0.043$ & 0.004 & 0.013 & 0.01 & 0.005 & 0.003 & 0.001 & 0.001 & 0.003 & 0.0 \\
$[3.0, 4.0]$ & $-0.074$ & 0.005 & 0.021 & 0.009 & 0.01 & 0.005 & 0.001 & 0.003 & 0.002 & 0.014 \\
$[4.0, 5.0]$ & $-0.032$ & 0.005 & 0.026 & 0.017 & 0.015 & 0.008 & 0.004 & 0.007 & 0.002 & 0.001 \\
$[5.0, 6.0]$ & $-0.026$ & 0.006 & 0.047 & 0.039 & 0.014 & 0.012 & 0.009 & 0.009 & 0.002 & 0.009 \\
$[6.0, 8.0]$ & $-0.039$ & 0.007 & 0.049 & 0.039 & 0.013 & 0.012 & 0.013 & 0.009 & 0.002 & 0.017 \\

\hline
$q_\perp$ [GeV] & $\langle \cos(2\phi) \rangle$ & Stat. & Tot. & Model & HFS(jet) & HFS(other) & HFS($\phi$) & Lepton(E) & Lepton($\phi$) & QED \\
\hline

$[0.0, 1.0]$ & 0.013 & 0.001 & 0.005 & 0.004 & 0.001 & 0.001 & 0.0 & 0.0 & 0.001 & 0.001 \\
$[1.0, 2.0]$ & 0.024 & 0.002 & 0.005 & 0.003 & 0.0 & 0.003 & 0.0 & 0.001 & 0.001 & 0.0 \\
$[2.0, 3.0]$ & $-0.003$ & 0.003 & 0.044 & 0.043 & 0.001 & 0.005 & 0.0 & 0.001 & 0.002 & 0.006 \\
$[3.0, 4.0]$ & $-0.01$ & 0.003 & 0.056 & 0.055 & 0.001 & 0.004 & 0.001 & 0.003 & 0.003 & 0.006 \\
$[4.0, 5.0]$ & $-0.039$ & 0.004 & 0.066 & 0.065 & 0.002 & 0.004 & 0.001 & 0.004 & 0.001 & 0.007 \\
$[5.0, 6.0]$ & $-0.049$ & 0.005 & 0.058 & 0.057 & 0.002 & 0.002 & 0.004 & 0.004 & 0.0 & 0.011 \\
$[6.0, 8.0]$ & $-0.048$ & 0.007 & 0.069 & 0.062 & 0.001 & 0.002 & 0.0 & 0.006 & 0.001 & 0.027 \\

\hline
$q_\perp$ [GeV] & $\langle \cos(3\phi) \rangle$ & Stat. & Tot. & Model & HFS(jet) & HFS(other) & HFS($\phi$) & Lepton(E) & Lepton($\phi$) & QED \\
\hline

$[0.0, 1.0]$ & 0.002 & 0.0 & 0.001 & 0.0 & 0.0 & 0.0 & 0.0 & 0.0 & 0.0 & 0.0 \\
$[1.0, 2.0]$ & 0.006 & 0.0 & 0.003 & 0.003 & 0.0 & 0.001 & 0.0 & 0.001 & 0.0 & 0.001 \\
$[2.0, 3.0]$ & $-0.002$ & 0.001 & 0.019 & 0.019 & 0.001 & 0.001 & 0.0 & 0.0 & 0.0 & 0.001 \\
$[3.0, 4.0]$ & $-0.029$ & 0.001 & 0.045 & 0.044 & 0.001 & 0.002 & 0.001 & 0.002 & 0.002 & $0.007$ \\
$[4.0, 5.0]$ & $-0.044$ & 0.001 & 0.055 & 0.054 & 0.002 & 0.002 & 0.001 & 0.002 & 0.004 & 0.013 \\
$[5.0, 6.0]$ & $-0.053$ & 0.002 & 0.056 & 0.051 & 0.003 & 0.003 & 0.003 & 0.002 & 0.003 & 0.022 \\
$[6.0, 8.0]$ & $-0.052$ & 0.002 & 0.046 & 0.045 & 0.002 & 0.004 & 0.003 & 0.002 & 0.002 & 0.001 \\
\hline

\end{tabular}
\label{tab:main_results}
\end{table}

\section{Conclusions}\label{sec:conclusions}
Results in this work were obtained using an unbinned unfolding deploying modern machine learning methods. The unfolding procedure was repeated from ~\cite{2108.12376,H1prelim-22-031}, and used to construct the observables presented in this work: the averages of the first three harmonics of the asymmetry angle $\phi$. This measurement depends on the unfolded moments of measured distributions, and would be impossible using traditional binned unfolding methods. 

A measurement of the azimuthal angular lepton-jet asymmetry in positron-proton collisions is presented. The measurement is shown as a function of the transverse momentum vector sum, $\qperp$. This observable is sensitive to initial and final state soft gluon radiation. This study also has very important implications for asymmetry studies, both at HERA and at the future EIC, where the contribution of soft gluon radiation that is not reconstructed as part of the jet, can become a substantial background to measurements that aim to explore the internal structure of the proton. For each harmonic, the data were tested for consistency with 0.0. It was found that all three distributions yield very low p-values of approximately 0, so no distribution was fully consistent with 0. The effect is quite small, however, with most points clearly consistent with 0.0 for $\qperp > 2$ GeV.

There is reasonably good agreement between the measurement and DIS event generators DJANGOH and RAPGAP. Both the nominal Pythia and tunes with alternative parton showers show larger disagreements with data. SHERPA at NLO and LO both show good agreement with the data at small $\qperp$, with the greatest difference in the first harmonic at larger $\qperp$. Similarly, both TXZ pQCD calculations show very good agreement with the data overall, but begin to deviate from the data for the first harmonic in the last $\qperp$ bin of the prediction, 3.0 GeV $< \qperp < 4.0$ GeV. Lastly, the HXYZ (TMD) prediction agrees with the data at small $\qperp$, but is not compatible with the data for $\qperp >$ 3.0 GeV in all three harmonics. These results could indicate that the condition of $\Pperp \gg \qperp$ begins to be compromised around $\qperp \approx 3.0$ GeV. As noted in Sec. \ref{sec:results}, the data are unable to differentiate between the TXZ(GBW) model, which incorporates saturation effects, and the TXZ(CT18A) model, which lacks these effects. Future eA collision measurements at the EIC or targeted HERA measurements at lower $Q^2$ values may be able to resolve this distinction.

A measurement of the asymmetry angle using jets with a resolution parameter of R=0.4 would be a very logical follow up study. As the jet cone is smaller, less soft gluon radiation would be reconstructed as part of the final state jet, and should therefore yield a larger measured asymmetry. Additionally, while the results show a small signal in the second and third harmonics for this work, potential next steps could be to measure the asymmetry angle in dijet and multi-jet DIS events, where the contribution of soft gluon radiation in the higher order harmonics is predicted to be much larger.

\section{Acknowledgements}
We would like to thank Bo-Wen Xiao, Feng Yuan, and Kyle Lee for helpful discussions. We would also like to thank Xuan-Bo Tong,  and Yuan-Yuan Zhang, for their calculations of the asymmetry angle, specifically carried out for HERA kinematics.
We are grateful to the HERA machine group whose outstanding efforts have made this experiment possible. We thank the engineers and technicians for their work in constructing and maintaining the H1 detector, our funding agencies for financial support, the DESY technical staff for continual assistance and the DESY directorate for support and for the hospitality which they extend to the non-DESY members of the collaboration.
We express our thanks to all those involved in securing not only the H1 data butq also the software and working environment for long term use, allowing the unique H1 data set to continue to be explored. The transfer from experiment specific to central resources with long term support, including both storage and batch systems, has also been crucial to this enterprise. We therefore also acknowledge the role played by DESY-IT and all people involved during this transition and their future role in the years to come.

L. Favart, T. Janssen, R. Roosen, P. Van Mechelen supported by FNRS-FWO-Vlaanderen, IISN-IIKW and IWT and by Interuniversity Attraction Poles Programme, Belgian Science Policy.
P.M. Jacobs, V.M. Mikuni, B. Nachman, F. Torales Acosta, H.T. Klest, A. Deshpande, A. Drees, C. Gal, S. Park, B.A. Schmookler, C. Sun, G. Tustin, S.H. Lee supported by the U.S. DOE Office of Science.
A. Buniatyan, P.R. Newman, P.D. Thompson, D.P.C. Sankey, R.C.W. Henderson, J.B. Dainton, T. Greenshaw, M. Klein, P. Kostka, J. Kretzschmar, S.J. Maxfield, A. Mehta, G.D. Patel, M.P.J. Landon, E. Rizvi, D. Traynor supported by the UK Science and Technology Facilities Council, and formerly by the UK Particle Physics and Astronomy Research Council.
M. Rotaru supported by the Romanian National Authority for Scientific Research under the contract PN 09370101.
D. Wegener, F. Huber, V. Radescu, M. Sauter, A. Schöning supported by the Bundesministerium für Bildung und Forschung, FRG, under contract numbers 05H09GUF, 05H09VHC, 05H09VHF, 05H16PEA.
L. Goerlich, S. Mikocki, G. Nowak, P. Sopicki partially supported by Polish Ministry of Science and Higher Education, grant DPN/N168/DESY/2009.
I. Picuric, N. Raicevic partially supported by Ministry of Science of Montenegro, no. 05-1/3-3352.
J. Cvach, J. Hladký, P. Reimer, T. Sykora, A. Valkárová, J. Žáček, R. Žlebčík supported by the Ministry of Education of the Czech Republic under the project INGO-LG14033.
K.B. Cantun Avila, J.G. Contreras supported by CONACYT, México, grant 48778-F.
C. Grab, T. Hreus, K. M\"uller, P. Robmann, U. Straumann supported by the Swiss National Science Foundation.

\begin{flushleft}
\bibliographystyle{desy24-200}

\bibliography{desy24-200}

\end{flushleft}

\end{document}